\documentclass[aps,prb,twocolumn,superscriptaddress,longbibliography]{revtex4-1}

\usepackage{hyperref} 
\usepackage{graphicx}
\usepackage{amsmath}
\usepackage{amssymb} 

\usepackage{hyperref}
\usepackage{float}
\usepackage[dvipsnames,svgnames,x11names,hyperref]{xcolor}
\DeclareGraphicsExtensions{.png,.jpg,.eps}
\usepackage{xcolor}

\renewcommand{\H} {\mathcal{H}}

\begin{document}

\author{Maryam Khosravian}
\affiliation{Department of Applied Physics, Aalto University, 00076 Aalto, Espoo, Finland}

\author{J. L. Lado}
\affiliation{Department of Applied Physics, Aalto University, 00076 Aalto, Espoo, Finland}

\title{Quasiperiodic criticality and spin-triplet superconductivity in
superconductor-antiferromagnet moire patterns
}

\begin{abstract}
Quasiperiodicity has long been known to be a potential platform to explore exotic phenomena,
realizing an intricate middle point between ordered solids and disordered matter. In particular,
quasiperiodic structures are promising playgrounds to engineer critical wavefunctions, a
powerful starting point to engineer exotic correlated states. Here we show
that
systems hosting a quasiperiodic
modulation of antiferromagnetism and spin-singlet superconductivity,
as realized by atomic chains in twisted van der Waals materials,
host a localization-delocalization transition
as a function of the coupling strength. 
Associated with this transition, we demonstrate
the emergence of a robust quasiperiodic
critical point for arbitrary incommensurate potentials, that
appears for generic relative weights of the spin-singlet
superconductivity and antiferromagnetism.
We show that inclusion
of residual electronic interactions 
leads to an emergent spin-triplet
superconducting state, that gets dramatically enhanced at the vicinity of the
quasiperiodic critical point.
Our results put forward
quasiperiodicity as a powerful knob to engineer robust superconducting states, providing
an alternative pathway
towards artificially designed unconventional superconductors.
\end{abstract}

\date{\today}

\maketitle
\section{Introduction}

Unconventional superconductivity \cite{RevModPhys.63.239,PhysRevLett.104.067001} encompasses one of the most exotic states found in quantum materials.
In particular, recent interest in topological superconductors
hosting Majorana states has been boosted
by their potential for topological
quantum computing \cite{PhysRevX.6.031016,Alicea2011}. However, unconventional superconductivity and,
in particular, spin-triplet superconductivity remains a highly elusive state in natural compounds \cite{PhysRevLett.84.5616,PhysRevB.70.014511,PhysRevB.97.104513,PhysRevB.77.064508,PhysRevLett.118.147001},
with a few exceptions such as doped Bi$_2$Se$_3$ \cite{Matano2016,Yonezawa2016,PhysRevB.90.100509,PhysRevB.94.180504}, UTe$_2$ \cite{Jiao2020}, and UPt$_3$ \cite{Avers2020}. Whereas
several compounds have been identified as a potential candidate for spin-triplet superconductivity \cite{PhysRevB.101.140503,PhysRevLett.84.1595,PhysRevB.75.020510,PhysRevB.100.035203,PhysRevLett.80.3129,PhysRevLett.89.037002}, finding generic mechanisms for its engineering still remains a challenge.
To date, a
highly successful strategy for engineering spin-triplet superconductors
consists of focusing on materials with potential coexisting magnetic and superconducting orders \cite{Alicea2012,Beenakker2013}.
This procedure has been heavily exploited for the engineering of Majorana bound states with a variety of platforms \cite{RevModPhys.83.1057,PhysRevLett.105.077001,PhysRevLett.105.177002,PhysRevLett.104.040502,PhysRevLett.100.096407,Feldman2016,Zhang2018,2020arXiv200202141K}. 
Most of these schemes have relied on engineering periodic structures with competing orders while its study in non-periodic systems has remained relatively unexplored \cite{PhysRevB.88.054204,PhysRevLett.110.146404}. In stark contrast,
the study of conventional superconductivity in disordered \cite{Anderson1959,PhysRevB.82.014509,PhysRevB.93.045111} and quasiperiodic systems\cite{PhysRevB.34.4390,PhysRevB.95.024509,PhysRevB.35.2494,PhysRevB.38.7436} 
has a long history, especially in demonstrating potential
critical advantages of non-periodicity for engineering robust superconducting states.

\begin{figure}[t!]
\centering
\includegraphics[width=\linewidth]{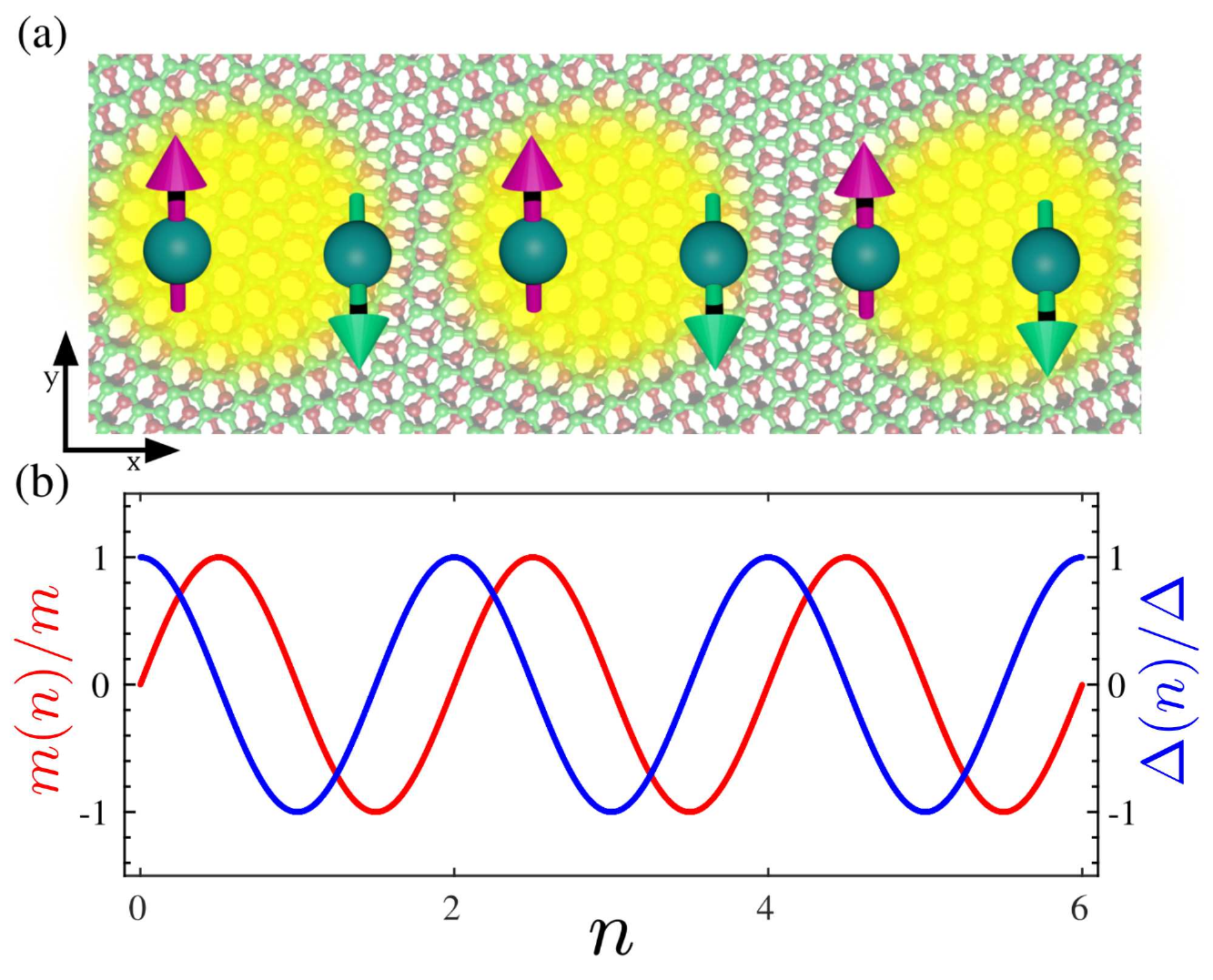}
\caption{
(a) Engineered antiferromagnetic atomic chain
on top of a superconducting twisted graphene
bilayer.
The twisted system shows a
spatially modulated
superconducting state (yellow),
that coexists
with the spatially modulated
antiferromagnetism of the chain.
Panel (b)
shows the modulation of
the superconducting 
$\Delta(n) = \Delta \cos{(\Omega n)}$
and
antiferromagnetic
$m(n) = m \sin{(\Omega n)}$
order parameters
along the 
chain direction,
associated to the Hamiltonian
of Eq. \ref{eq:h}.}
\label{fig:sketch}
\end{figure}

Quasiperiodic patterns display a never-repeating arrangement of elements\cite{PhysRevLett.53.1951, PhysRevLett.53.2477},
yet they host a long-range order. Due to the lack of conventional periodicity, standard band-structure arguments no longer hold, and their electronic structure exhibits a notably rich behavior \cite{Jagannathan2007}, such as the presence of confined states \cite{Kohmoto1986, Arai1988, Rieth1995}, fractal dimensions \cite{Sutherland1986, Tsunetsugu1986, Tokihiro1988}, pseudogap in the density of states \cite{Fujiwara1989, Fujiwara1991,Ishikawa2017,Jazbec2014}, and unconventional conduction properties \cite{Pierce1993, TramblydeLaissardiere1994, Timusk2013, deLaissardiere2014}. 
More importantly, the incommensurate structure of quasicrystals has prominent effects on the electron
eigenstates. Incommensurate potentials give rise to
electronic wave functions extended, localized or critical \cite{Siebesma1987,aubry1980analyticity, Devakul2017,Su2018}, 
and ultimately can host topological states of matter fully associated to the quasiperiodicity \cite{PhysRevLett.109.106402,PhysRevLett.109.116404}.

Here we demonstrate that quasiperiodic patterns
arising from a combination
of spin-singlet superconductivity
and antiferromagnetism provide
a powerful platform to engineer spin-triplet superconductivity. In particular,
we show that this antiferromagnetic-superconductor
pattern hosts a localization-delocalization
phase transition, with an associated quasiperiodic critical
point with multifractal wavefunctions.
We further show that upon inclusion of residual
interactions,
a spin-triplet superconducting state
emerges that gets dramatically enhanced
at the proximity of the
localization-delocalization critical point.
Our results show that magnetic-superconducting
quasiperiodic patterns, as those found
in atomic chains in twisted van der Waals materials,
provide a new mechanism to engineer
unconventional superconducting states
by exploiting quasiperiodic criticality.

Our manuscript is organized as follows. Sec. \ref{sec:critical} introduces
a realization of our model, and we
show the emergence
of a critical point in quasiperiodic superconductor-antiferromagnet patterns,
separating the extended and localized regime.
In Sec. \ref{sec:interactions}, we show that interactions give rise to a spin-triplet
superconducting state, and we analyze its dependence
with respect to the details of the quasiperiodic modulation. In Sec. \ref{sec:per}, we demonstrate
the robustness of the interaction
induced spin-triplet
superconducting state
with respect to perturbations
in the quasiperiodic Hamiltonian. Finally, in
Sec \ref{sec:con}, we summarize our conclusions.

\section{Antiferromagnet-superconductor quasiperiodic criticality}
\label{sec:critical}

The system that we will study combines a spatially modulated antiferromagnetism and superconductivity,
as shown in Fig. \ref{fig:sketch}ab.
This type of spatially modulated parameters appear in generic twisted two-dimensional materials
that combine superconductivity and magnetism \cite{2020arXiv200202141K}. Here we will focus on a specific case
in which the system is purely one dimensional. This situation can be realized by taking a
twisted graphene multilayer in a superconducting state \cite{Cao2018,Yankowitz1059,Lu2019,PhysRevB.99.121407,PhysRevLett.121.087001,PhysRevB.101.060505,PhysRevB.98.220504}, 
whose superfluid density follows the
modulation of the moire pattern, and depositing an array of
ad-atoms on top of it \cite{Gonzalez-Herrero437,Toskovic2016,Kezilebieke2019,Loth2012,Liebhaber2019,PhysRevLett.115.197204,CortsdelRo2020,RevModPhys.91.041001} (Fig. \ref{fig:sketch}a). The ad-atoms
will have a long-range antiferromagnetic order stemming from the graphene RKKY interaction \cite{PhysRevB.84.115119}, leading to a one-dimensional antiferromagnetic state \cite{PhysRevB.82.045407,PhysRevResearch.1.033173,PhysRevResearch.1.033009}. Both electronic orders will have a modulation
following the moire pattern, effectively realizing to a one dimensional
model with modulated antiferromagnetism and superconductivity \cite{Brihuega2017,PhysRevMaterials.3.084003,PhysRevB.99.245118,PhysRevResearch.2.033357}.

We describe this system combining a modulated superconducting and antiferromagnetic
exchange by the following effective Hamiltonian \cite{PhysRevResearch.2.023347,PhysRevLett.121.037002,PhysRevB.100.125411,PhysRevX.5.041042}:

\begin{eqnarray}\label{e37}
\H&=&t \sum_{n,s} c^{\dagger}_{n, s}c_{n+1,s} +h.c.
\nonumber \\
&+&m \sum_{n,s,s'}\sigma_z^{s,s'} (-1)^n\sin(\Omega n)c^{\dagger}_{n,s}c_{n,s'} \nonumber \\
&+& \Delta\sum_{n} \cos(\Omega n) c^{\dagger}_{n,\uparrow}c^{\dagger}_{n,\downarrow}+h.c.
\label{eq:h}
\end{eqnarray}
where $c^{\dagger}_{i,s}$ ($c_{i,s}$) denotes the
fermionic creation (annihilation) operator for site $n$
and spin $s$,
and $\sigma_z$ is the spin Pauli matrix. 
The first term denotes the kinetic energy of the system,
the second term denotes the spatially modulated antiferromagnetism, and the third term
corresponds to the modulated superconductivity. The parameters $\Delta$ and $m$ are responsible for the
strength of the modulation corresponding to the antiferromagnetism and superconductivity,
respectively, and $\Omega$ 
is the wavelength of the modulation. The model of Eq. \ref{eq:h} assumes that the proximity-induced
superconducting state
will be stronger when the magnetism is weaker (Fig. \ref{fig:sketch}b), as often happens for spin-singlet
superconductivity. 
In the following, the s-wave superconducting order will be taken as a parameter of the model
as stemming from proximity. When interactions are included, a different superconducting
order can appear on top of the proximitized one. This additional superconducting order
arising when interactions are included will be called interaction-induced superconducting order.
For convenience, we will parameterize the superconducting and antiferromagnetic strength as
$m = \lambda \sin{\alpha}$ and $\Delta = \lambda \cos{\alpha}$, so that the net strength of the quasiperiodic modulation
can be defined by the parameter $\lambda = \sqrt{m^2 + \Delta^2}$.

\begin{figure}[t!]
\centering
\includegraphics[width=\linewidth]{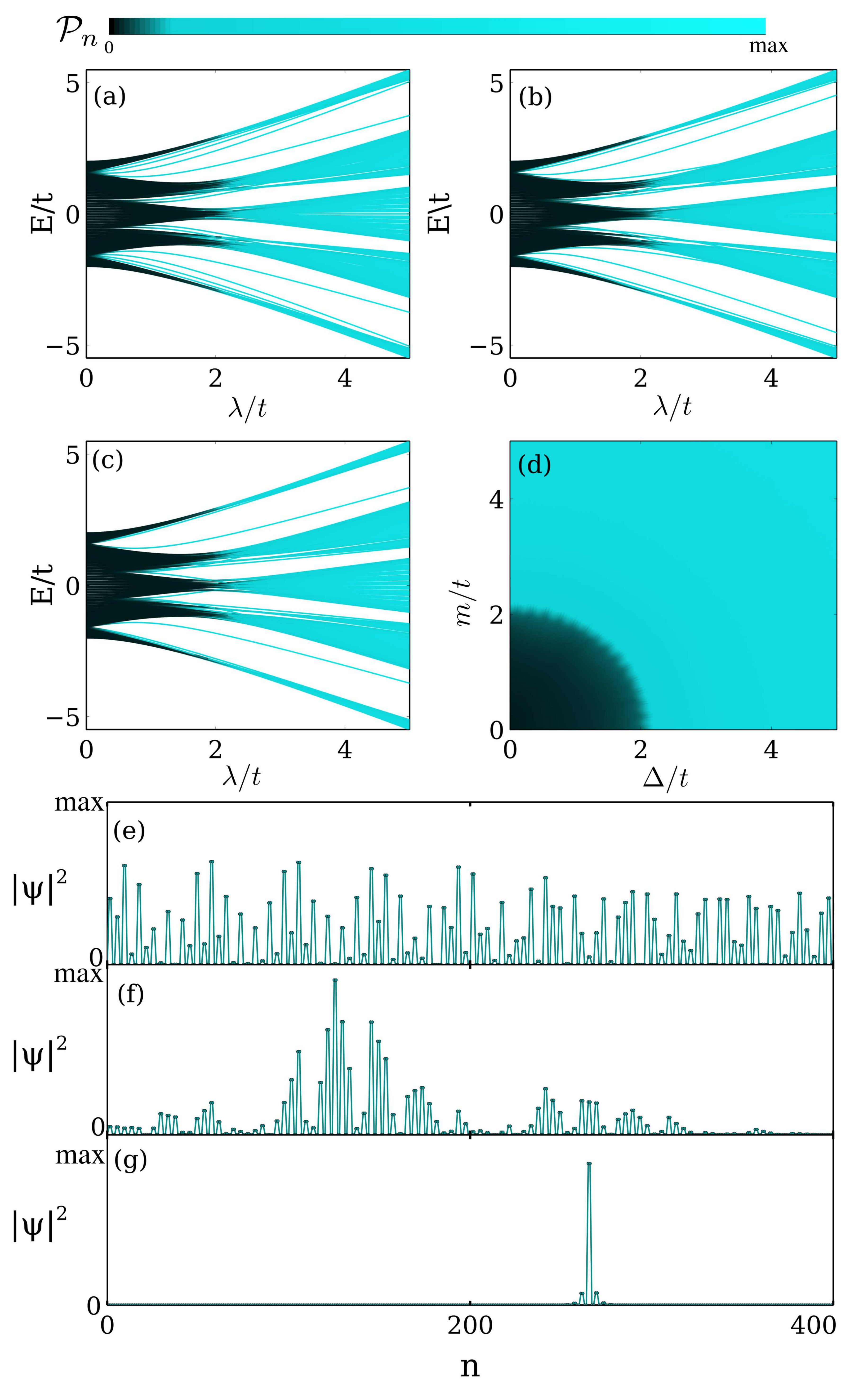}
\caption{(a,b,c) IPR for every state as a function of the 
modulation strength, with $\alpha=\pi/3$ in (a),
$\alpha=\pi/4$ in (b)
and 
$\alpha=\pi/6$ in (c).
It is observed that the
the localization transition takes place for $\lambda=2t$
in all instances. (d) IPR averaged
over all the states as a function
of $m$ and $\Delta$, showing
a phase boundary following $\Delta^2 + m^2 = 4t^2$. Panels
(e-g) show the
spatial distribution of the wave function in the extended (e), critical (f), and localized regime (g).}
\label{fig:ipr}
\end{figure}

For irrational values of $\Omega/(2\pi)$, the model of Eq. \ref{eq:h}
lacks translational symmetry and thus does not accept a description in terms of Bloch states.
As a result, the eigenstates of this Hamiltonian are not guaranteed to be extended states,
as the Hamiltonian is inherently non-periodic.
Whereas random disorder creates localization at arbitrarily small coupling constants in one-dimension \cite{PhysRev.109.1492}, quasiperiodic patterns are known to give rise to a localization transition at finite coupling constant \cite{PhysRevLett.50.1870,aubry1980analyticity}.
In particular, it is worth noting that for $\Delta=0$, our model is mathematically equivalent
to the Aubry-Andre-Harper model \cite{aubry1980analyticity}. Therefore, in the limit of $\Delta=0$, the
previous model will have a localization transition at $m=2t$, so that
for $m<2t$ all the states will be extended and for $m>2t$ all the states will be localized. 
As we show below, the generalized
model of Eq. \ref{eq:h} with $\Delta\ne 0$ 
shares many of the characteristics of the AAH model, in particular a critical transition
at finite coupling constant.

We now address the localization-delocalization transition
in the previous model. To determine the extended and localized nature of the states, we 
compute the inverse participation ratio (IPR) of each eigenstate $\Psi_n$ as
\begin{eqnarray}\label{e37}
\mathcal{P}_{n}=\sum_{i}|\Psi_{n}(i)|^4
\end{eqnarray}
where $i$ runs over all the components of each eigenstate.
For localized states whose wavefunction spans a certain number of sites $L$, for a system of size $N$, the value
of the IPR is a finite non-zero number. In stark contrast, for extended states,
the value of the IPR scales as $1/N$ 
becomes zero in the thermodynamic limit.

Let us now explore the model of Eq. \ref{eq:h}, and in particular, analyze how the localization of the states evolve as we increase the strength of the quasiperiodic modulation $\lambda$.
We show in Fig. \ref{fig:ipr}a the evolution of the
IPR for the different eigenstates, as a function of the modulation strength $\lambda$
for $\alpha=\pi/3$, which corresponds to taking $\Delta/m \approx 0.57$. 
In particular, as shown in Fig. \ref{fig:ipr}abc, all the states remain extended
up to $\lambda=2t$ is reached, at which point all become localized.
Note that the single in-gap modes that remain all the time localized correspond
to topological edge states \cite{PhysRevLett.109.106402,PhysRevLett.109.116404}.
This can be systematically studied by looking at the average IPR of the states
as a function of $\Delta$ and $m$. This is shown in
Fig. \ref{fig:ipr}d, where it can be seen
that a boundary with the functional form $\Delta^2 + m^2 = \lambda^2 = 4t^2$
separates the localized region from the extended region.
The transition's independence with respect to $\alpha$ can be
rationalized from a low energy model. In particular,
for a tight-binding chain with antiferromagnetism and superconductivity, the low energy
model consists of a four-component Dirac equation \cite{PhysRevX.5.041042,PhysRevB.100.125411,PhysRevLett.121.037002,PhysRevResearch.2.023347,2020arXiv201106990L}. The antiferromagnetic $m$ and
superconducting $\Delta$ order parameters enter in this low energy model as
two inequivalent masses $m(x)$ and $\Delta(x)$ in the Dirac equation. 
Thus, from the low energy perspective,
the quasiperiodic model can be understood as a Dirac equation in which two mass terms
$m(x)$ and $\Delta(x)$
are modulated in space. In particular, performing a local 
spinor rotation of the Dirac model, we reach an effective model
with a single Dirac equation with a mass term $\chi(x) = \sqrt{m(x)^2 + \Delta(x)^2}$.
This spinor rotation does not change the localized or delocalized nature of the
eigenstates.
By definition of $m(x)\sim\lambda \cos{\alpha}$
and $m(x)\sim\lambda \sin{\alpha}$, 
$\chi(x)$ is independent of $\alpha$, and therefore
the localization-delocalization becomes independent of $\alpha$.

At the previous phase boundary separating extended from localized states,
wavefunctions with critical behavior emerge.
The different nature of the extended, localized, and critical states
can be easily observed by plotting individual wavefunctions.
In particular, we show in Fig. \ref{fig:ipr}efg the wavefunction closest to charge
neutrality for an extended (Fig. \ref{fig:ipr}e), critical (Fig. \ref{fig:ipr}f)
and localized (Fig. \ref{fig:ipr}g) regime. Whereas the extended wavefunctions
span over the whole system (Fig. \ref{fig:ipr}e), localized wavefunctions are strongly localized
in a few lattice sites (Fig. \ref{fig:ipr}g).
The critical wavefunction of Fig. \ref{fig:ipr}f is characterized by
multifractal revivals \cite{PhysRevA.87.023625,Goblot2020,PhysRevB.42.8244,PhysRevB.93.104504,PhysRevB.62.15569,Castellani1986,Falko1995}. This will become especially 
important in the next section, as the multifractal behavior of the
states will substantially increase the impact of interactions
in the system.

\section {Interaction-driven spin-triplet superconductivity}
\label{sec:interactions}

\begin{figure}[t!]
\centering
\includegraphics[width=0.8\linewidth]{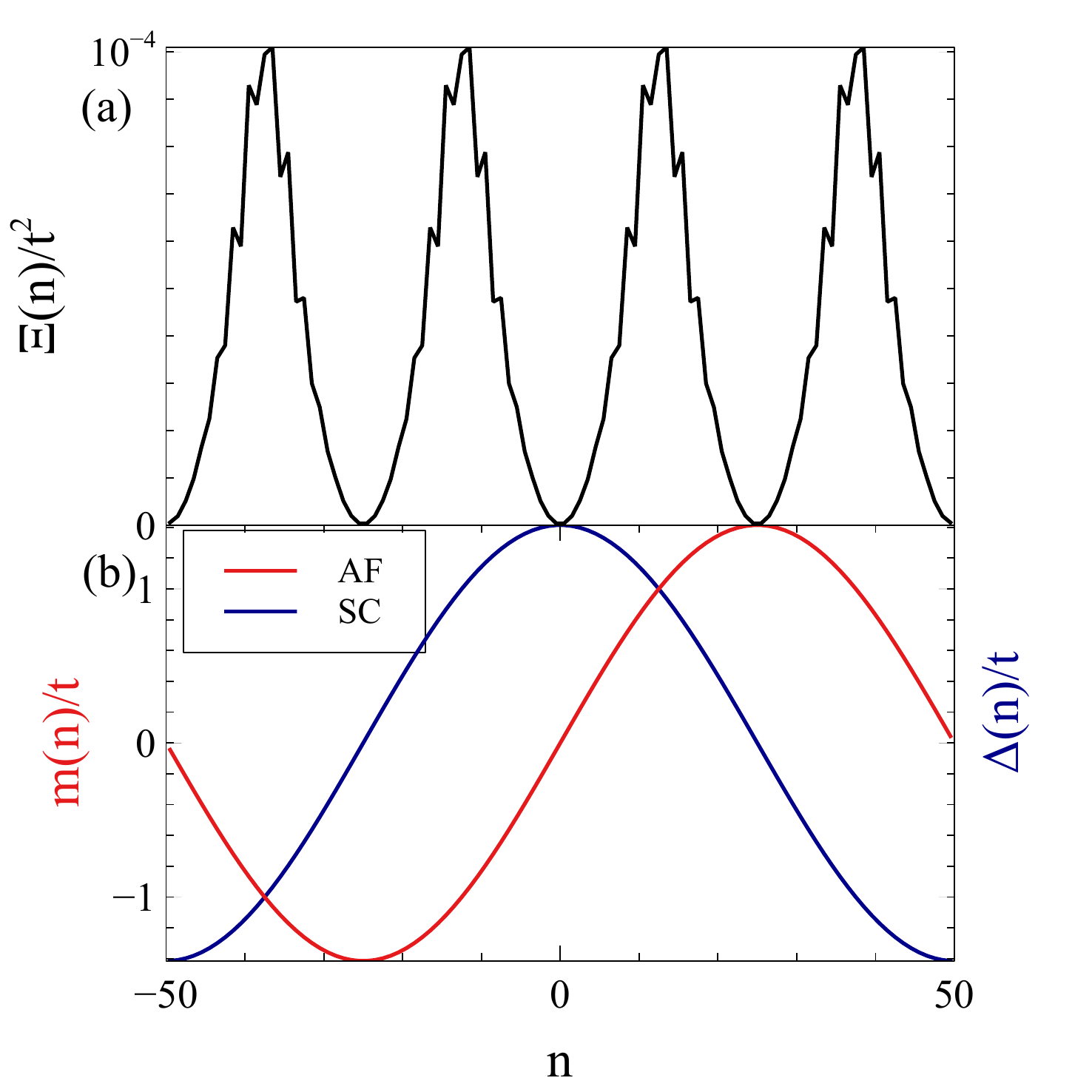}
\caption{Selfconsistent interaction-induced
spin-triplet superconducting order as a function
of the position
(a), and spatial profiles of the spin-singlet
superconductivity $\Delta(n) = \Delta \cos{(\Omega n)}$ 
and antiferromagnetism
$m(n) = m \sin{(\Omega n)}$
(b).
It is observed that the spin-triplet component
is maximal in the regions where the spin-singlet
superconductivity
and antiferromagnetism coexist.}
\label{fig:spatial}
\end{figure}

Let us now move on to consider the impact of interactions in the previous quasiperiodic
system. 
In particular, 
we will show that the inclusion of interactions will lead
to spin-triplet superconductivity,
where the criticality driven by the quasiperiodic pattern can be 
used as a knob to enhance superconducting order parameter close to
the critical point.
Local interactions of the form
$    \H_{\text{int}} \sim \sum_n c^\dagger_{n,\uparrow} c_{n,\uparrow} c^\dagger_{n,\downarrow} c_{n,\downarrow} $
are already accounted for in the stagger antiferromagnet and superconducting terms of the Hamiltonian.
Therefore, we will now consider the effect of residual non-local
interactions, in particular nearest neighbor density-density interactions. 
For that sake, we will
now include an interaction term in our Hamiltonian of the form

\begin{equation}
\label{eq:hint}
    \H_{V} = -V
    \sum_n 
    \left ( 
    \sum_{s}  c^\dagger_{n,s} c_{n,s} \right )
        \left ( 
        \sum_{s'}  c^\dagger_{n+1,s'} c_{n+1,s'} \right )
\end{equation}
where $V$ controls the strength of the nearest-neighbor attractive
interaction. In the the following, we will solve the previous Hamiltonian
by a mean-field decoupling.
Before moving forward with out results,
it is first worth to note that
quantum many-body methods such as bosonization and density-matrix 
renormalization group\cite{PhysRevLett.69.2863} (DMRG) are used to obtain accurate results
in interacting one dimensional systems \cite{PhysRevLett.83.3908,PhysRevB.65.115114}. 
These methods fully capture quantum fluctuations, which are missing in mean-field approaches. 
Mean-field theory methods are used to approximately study interacting quasiperiodic models in
one-dimension, providing qualitatively correct results that can later be refined with more advanced
quantum many-body methods. In particular, mean-field methods in one-dimension have been benchmarked 
with density-matrix renormalization group calculations for Aubry-Andre-Harper
models \cite{PhysRevB.89.161106,PhysRevB.91.104203}, finding a good qualitative agreement between
mean-field and DMRG methods\cite{PhysRevB.89.161106,PhysRevB.91.104203}. Following those results, 
it is expected that results obtained with mean-field theory provide a qualitatively correct picture.
Specifically, a particular difference between the mean-field
and full many-body approaches, is that mean-field solutions would 
predict long-range order, whereas
exact solutions would rather show quasi long-range
order, with a power-law decay of correlations in the order
parameter \cite{PhysRevA.82.043613}.

Keeping the previous points in mind, we now move on to 
solve the previous interacting term Eq. \ref{eq:hint} using a mean-field approximation
$\H^{\text{MF}}_V$, including all the normal $\H^{\text{MF},h}_V$ and
anomalous contributions $\H^{\text{MF},s}_V,\H^{\text{MF},t}_V$. 
The previous mean-field decoupling will give rise to hopping renormalization
$\H^{\text{MF},h}_V$,
singlet superconductivity renormalization
$\H^{\text{MF},s}_V$
and most importantly, to potential spin-triplet superconducting order
$\H^{\text{MF},t}_V$. We
will focus in this last term, whose contribution to the mean-field Hamiltonian is of the form

\begin{equation}
    \H^{\text{MF},t}_V = 
    \sum_n \Delta^{ss'}_{n,n+1} c_{n,s} c_{n+1,s'} +
    \text{h.c.}
\end{equation}
where by definition of the fermionic
anticommutation relations
$\Delta^{s,s'}_{n,n+1} = - \Delta^{s',s}_{n+1,n}$.
It is worth to note that the interaction term we consider
can lead to both spin-singlet and spin-triplet
components. In our calculations, we have verified that the spin-singlet
component is zero in the cases under study. This can be intuitively understood
from the fact that the existence of local antiferromagnetism
is expected to quench spin-singlet instabilities.\cite{Anderson1959,PhysRevB.30.4000,PhysRevB.98.024501,Andersen2020}
Therefore, whereas spin-singlet instabilities may appear
from Eq. \ref{eq:hint}, the spin-triplet component
is the leading instability of the model under study.
In the following, we will focus in the spin-triplet component of the
superconducting order, that fulfills
$\Delta^{s,s'}_{n,n+1} = - \Delta^{s,s'}_{n+1,n}$. As the spin-triplet
component of the interaction induced superconducting state has several degrees of freedom, it is convenient
to define an spatially dependent d-vector $\vec d_{n,n+1}$ that parameterizes the spin-triplet superconducting
order as

\begin{eqnarray}
\Delta^{ss'}_{n,n+1}=
 \left ( 
	\begin{array}{c c }
     \Delta^{\uparrow \uparrow}_{n,n+1} & 
		\Delta^{\uparrow \downarrow}_{n,n+1} \\
		\Delta^{\downarrow \uparrow}_{n,n+1}& 
		\Delta^{\downarrow \downarrow}_{n,n+1} \\
 \end{array}
	\right)
	= 
	i \sigma_{y}(\Vec{d}_{n,n+1}.\vec{\sigma}),\nonumber \\
\end{eqnarray}
where $\vec \sigma$ are the spin-Pauli matrices. 

As a first step, it is interesting to look at the real-space distribution of the unconventional superconducting state. 
In particular we show in Fig. \ref{fig:spatial} the real-space distribution of the spin-triplet state
defined as

\begin{equation}
    \Xi(n) = \sum_j |\vec d_{n,n+j}|^2 = |\vec d_{n,n-1}|^2 + |\vec d_{n,n+1}|^2
    \label{eq:stri}
\end{equation}
we note that in the case of first neighbor interactions, the previous sum
includes only $j=-1$ and $j=+1$, yet in the presence of longer range interactions
other terms could be included.
We observe that the spin-triplet density follows the quasiperiodic pattern,
and that it becomes zero in regions only having antiferromagnetism or s-wave superconductivity
(Fig. \ref{fig:spatial}ab).
In particular, the value of the spin-triplet superconducting order becomes maximal
every time the s-wave superconductivity and antiferromagnetism coexist maximally
in absolute value, irrespective of their signs.
Interestingly, we find that such spin-triplet component is maximal right in the region where the
s-wave superconductivity and antiferromagnetism coexist in the same footing (Fig. \ref{fig:spatial}ab),
highlighting the key interplay of magnetism and superconductivity for
driving as spin-triplet superconducting state.
Moreover, it is interesting to examine the specific type of spin-triplet state
that the interactions
promote. In particular, we find that the $\vec d_{n,n+1}$ is always locked to the same direction
of the antiferromagnetism, which in terms of the superconducting order parameters is associated
to an interaction induced spin-triplet $\Delta^{\uparrow\downarrow}_{n,n+1}$ order parameter
for antiferromagnetism in the $z$-axis.

\begin{figure}[t!]
\centering
\includegraphics[width=\linewidth]{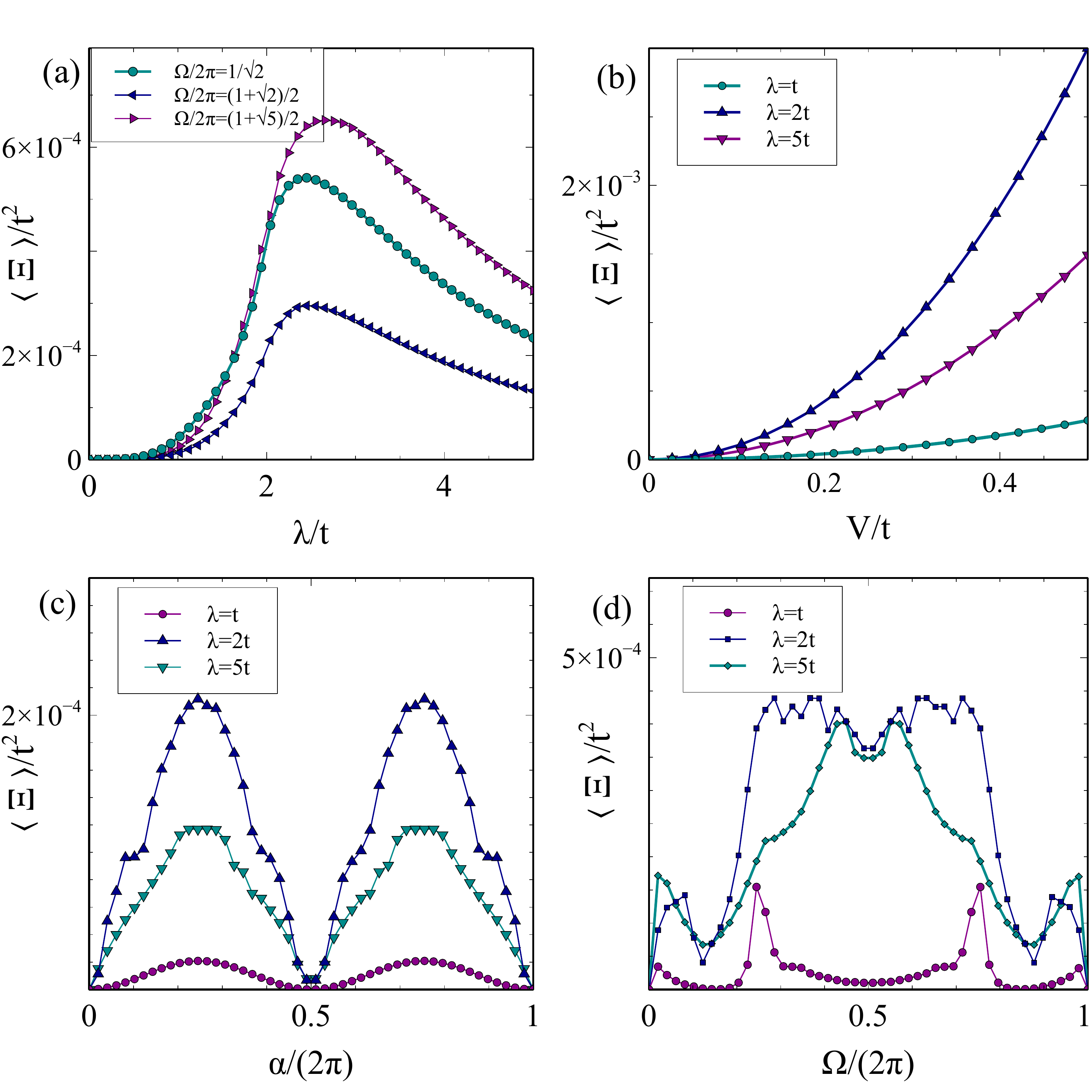}
\caption{Selfconsistent
interaction-induced spin-triplet superconductivity as a function of
the superconductor-antiferromagnet
modulation strength for $V=0.2t$, $\alpha=\pi/4$ (a), as a function of the interaction strength V for $\alpha=\pi/4$, $\Omega/(2\pi)=1/\sqrt{2}$ (b), 
as a function of the SC-AF angle $\alpha$ for $\Omega/(2\pi)=1/\sqrt{2}$, $V=0.2t$ (c), and as a function of
the quasiperiodic modulation frequency $\Omega$ for $V=0.2t$, $\alpha=\pi/4$ (d).
An enhancement of the interaction
induced spin-triplet superconducting state is observed, 
stemming from the interplay of antiferromagnetism
and superconductivity.
}
\label{fig:dos}
\label{fig:scfcrit}
\end{figure}

We now move on to examine the impact of the quasiperiodic criticality on the induced
spin-triplet state. As anticipated above, the critical behavior of the wavefunctions is
known to provide an effective mechanism for enhancing electronic instabilities \cite{PhysRevB.102.115108,RubioVerd2020,PhysRevB.88.195139,PhysRevLett.108.017002,Petrovi2016,2020arXiv200800503F}.
To verify this, we now compute the selfconsistent spin-triplet order parameter
of Eq. \ref{eq:stri}
averaged over all the sites
$\langle \Xi \rangle = \frac{1}{N}\sum_n \Xi (n)$
as a function of the modulation strength $\lambda$, as shown in
Fig. \ref{fig:scfcrit}a. It is observed that, as the modulation strength increases, the
induced spin triplet parameter grows, becoming maximal around the critical point and decreasing as
the system goes deeper into the localized regime. The enhancement associated
to the critical point can also be verified by computing the
induced spin-triplet order parameter as a function of the interaction strength $V$,
as shown in Fig. \ref{fig:scfcrit}b. It is seen that for all coupling constants, the
spin-triplet state is stronger close to the critical point
$\lambda=2t$, than deep into the extended ($\lambda=t$) or localized ($\lambda=5t$)
regime. The fact that the maximum does not exactly
appear at the critical point, but towards the localized region
can be rationalized as follows.
In the localized limit, a sizable superconducting stability of isolated electronic states
can happen at small coupling. In this localized regime, the absence of kinetic energy
for the localized modes yields a large expectation value of the
superconducting order that explains the existence of a sizable
spin-triplet SC in the localized regime, and a maximum that is not exactly at the critical point.
However, in this localized regime, the different states that become
ordered are spatially separated, which would prevent the existence of phase 
coherence between them \cite{PhysRevA.82.043613,PhysRevB.101.104509}. While those fluctuations of the phase coherence are not captured at the mean-field level, a calculation
that includes quantum fluctuations would show that the phase coherence in this localized regime
is very small \cite{PhysRevA.82.043613,PhysRevB.101.104509}. 
In contrast, for $\lambda < 2t$, the extended nature of the
states would give a superconducting state with a large phase coherence and superconducting stiffness.

We now move on to examine the impact of the two quasiperiodic modulations as parameterized
by $\alpha$. As we showed above, the extended-localized transition takes place for
$\lambda=2t$, and independently on the value of $\alpha$. This means that a critical
point appears independently on the relative strengths between $\Delta$ and $m$,
and it only depends on $\lambda = \sqrt{\Delta^2 + m^2}$. In stark contrast, the
emergence of a spin-triplet component due to interactions
turns out to be highly dependent on $\alpha$, as shown in Fig. \ref{fig:scfcrit}c. In
particular, we observe that when the system is purely antiferromagnetic 
or purely superconducting ($\alpha = 0,\pi/2$) the spin-triplet
component generated is exactly zero. In comparison, the spin-triplet
state is maximal for $\alpha=\pi/4$, that corresponds to having an equal weight on the
singlet superconducting and antiferromagnetic order parameters. This observation
emphasizes the importance of the coexistence of antiferromagnetism and
superconductivity for the emergence for the interaction induced
spin-triplet state.

We finally consider the impact of the spatial modulation
frequency $\Omega$ in the interaction induced
spin-triplet state. As shown in Fig. \ref{fig:scfcrit}d we observe that
the enhancement at the critical point happens for generic values
of the modulation frequency. In the limit of small $\Omega$, the system
is essentially formed by patches of superconductor and
antiferromagnet \cite{PhysRevResearch.2.023347,PhysRevLett.121.037002,PhysRevB.100.125411,PhysRevX.5.041042}, having a typical length on the order $\l \sim 1/\Omega$, and
thus the interaction between the superconducting and antiferromagnetic state
happens in a limited part of the system. In comparison, for
$\Omega \approx \pi/2$ there is a quick oscialltion between the two orders,
promoting a dense coexistence of antiferromagnetism and spin-singlet
superconductivity in the system. We observe that the interaction induced
spin-triplet component is especially strong in this regime (Fig. \ref{fig:scfcrit}d), reflecting
the key interplay between spin-singlet superconductivity and
antiferromagnetism for driving the unconventional
superconducting state.

\begin{figure}[t!]
\centering
\includegraphics[width=0.8\linewidth]{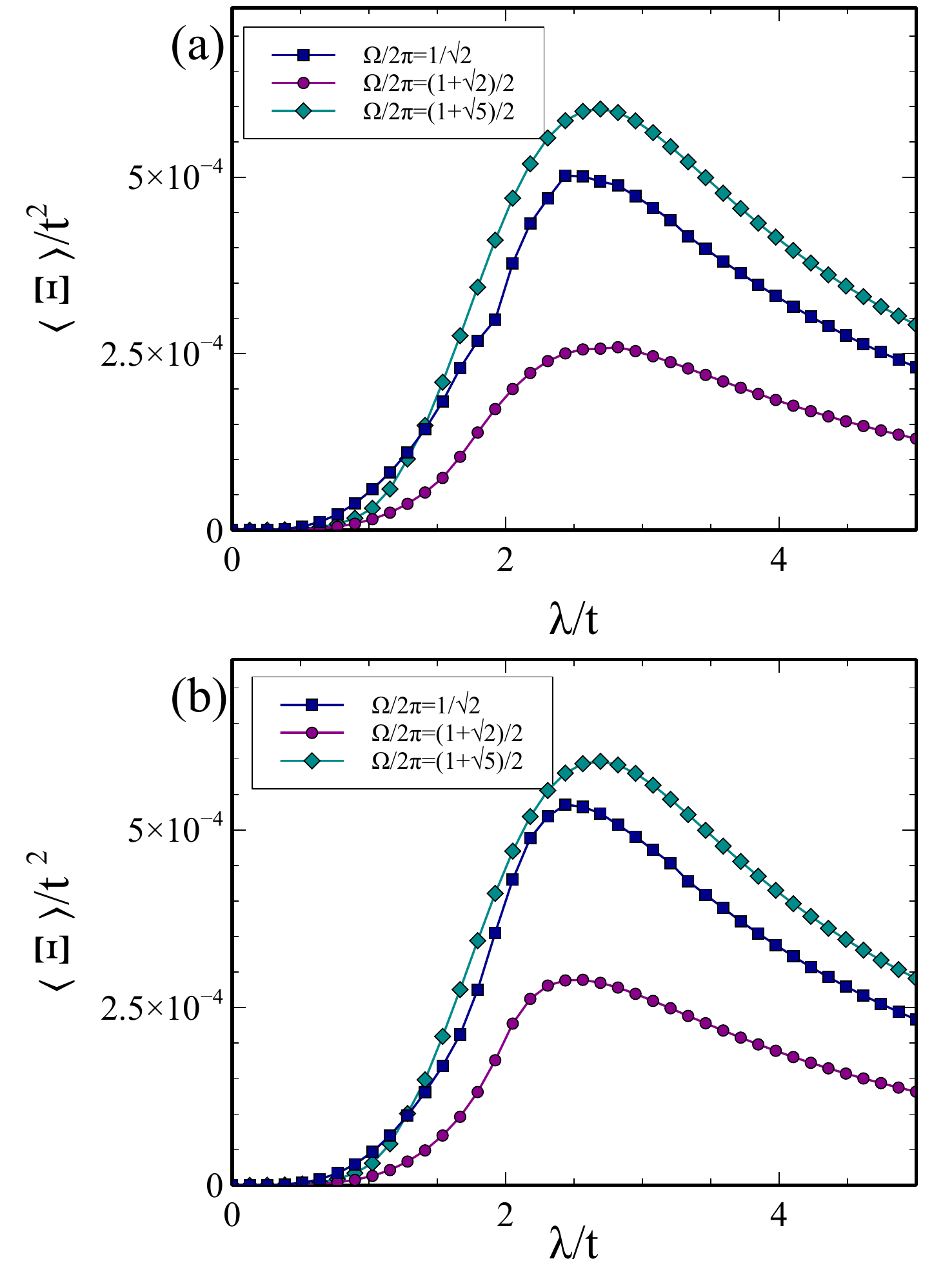}
\caption{Spin-triplet superconducting state
in the presence of second neighbor hopping for $\alpha=\pi/4$, 
	$V=0.2t$ (a) and random
disorder for $\alpha=\pi/4$, $V=0.2t$ 
(b). It is observed that the enhancement close to
the critical point survives
perturbations to
the quasiperiodic Hamiltonian of Eq. \ref{eq:h}.}
\label{fig:per}
\end{figure}

\section{Robustness to perturbations}
\label{sec:per}

In this section, we address the robustness of our phenomenology with respect to perturbations. In particular,
we will focus on the impact of next to nearest-neighbor hopping 
and Anderson disorder. The next to nearest-neighbor hopping breaks the bipartite
nature of the lattice, whereas the Anderson disorder
would drive the system to a localized state for all $\lambda$.
In particular, we obtain that the critically enhanced spin-triplet superconducting state
also happens with those additional perturbations, as elaborated below.
Second neighbor perturbations are expected to appear in a realization of the previous model and break the original
bipartite nature of the system. The previous term is included in our Hamiltonian by means of a perturbation of the form

\begin{equation}
    \H_{\text{NNN}} = t'
    \sum_{n,s}
    c^\dagger_{n,s} c_{n+2,s} + \text{h.c.}
\end{equation}
The results with this additional perturbation are shown in Fig. \ref{fig:per}a, where we took $t' =0.2t$.
It is observed that the enhancement of the interaction-induced spin-triplet state happens
in the presence of this additional perturbation. It is worth to note that in the presence of second
neighbor hoppings, the localization-delocalization transition becomes state dependent, and it will
no longer happen at $\lambda=2t$. Nevertheless, it is observed that the qualitative behavior
remains analogous to the idealized case with $t'=0$. This is especially important for potential
realizations of our model in twisted two-dimensional materials and cold atoms setups, as generically
these systems present small additional contributions to the Hamiltonian such as a second neighbor
hopping.

Next, we consider the impact of random disorder in the system, included as an onsite Anderson perturbation.

\begin{equation}
    \H_{\text{dis}} = 
    \sum_{n,s} \epsilon_n c^\dagger_{n,s} c_{n,s}
\end{equation}

where $\epsilon_n$ is a random number between $[-0.1,0.1]t$.
First, it is interesting to note that the inclusion of an arbitrarily small amount of disorder would
drive the extended states to a localized regime.
As a result, in the presence of disorder the localization-delocalization transition
as a function of $\lambda$ is completely destroyed, as the state becomes localized for all
$\lambda$. The disorder strength $\lambda$ will define a minimal localization
length for the system. As $\lambda$ is ramped up, the system will go from a localized
regime dominated by the disorder, to a regime in which the localization
is dominated by the quasiperiodic potential. Although the critical
point is washed out, the enhancement of the superconducting state
will still be visible at this quasiperiodic-disorder localization
crossover.
This is shown in Fig. \ref{fig:per}b, where it is seen that the spin-triplet
enhancement close to the former critical
point is still visible. This phenomenology shows
that even in experimental
setups that host small imperfections,
the enhancement of an unconventional
spin-triplet superconducting state
can be observed.

Finally, we comment on the prospects of extending the previous phenomenology
to two-dimensional systems. In our manuscript, we focused
on showing the existence of a localization-delocalization
transition for a quasiperiodic one-dimensional model. Right
at the transition, the existence of a critical point lead to
an enhancement of an interaction-induced instability,
that due to the nature of our system was a spin-triplet
superconducting instability. Interestingly, quasiperiodic
system in higher dimensions also show localization-delocalization
transitions, and associated critical points to them.\cite{PhysRevB.96.045138}
Enhanced symmetry broken orders induced
by interactions have been found in those quasiperiodic
two-dimensional models,\cite{PhysRevB.95.024509,PhysRevB.100.014510,PhysRevResearch.1.022002,PhysRevLett.125.017002,PhysRevB.102.224201,PhysRevB.102.115125} both in the
cases of spin-singlet
superconductivity\cite{PhysRevB.95.024509,PhysRevB.100.014510,PhysRevResearch.1.022002,PhysRevLett.125.017002} and magnetic order.\cite{PhysRevB.102.115125}
The previous phenomenology in two-dimensions
was also demonstrated for one-dimensional models, highlighting
that symmetry breaking enhanced by quasiperiodicity happens both in
one-dimensional and two-dimensional quasiperiodic models.
Ultimately the previous results suggest that extension to two-dimensions
of our model could lead to two-dimensional spin-triplet superconductivity.

\section{Conclusion}
\label{sec:con}
To summarize, 
we have demonstrated that antiferromagnet-superconductor moire patterns
show a critical point associated with a localization-delocalization
transition. We showed that the quasiperiodic
criticality happens
for arbitrary ratios between the superconducting and
antiferromagnetic order parameters, and that the critical
point is universally located in a curve defined by the
two order parameters. Upon inclusion of residual electronic interactions,
we demonstrated the emergence of an unconventional
spin-triplet state, whose $d$-vector is locked along the
antiferromagnetic spin direction. We showed that the emergence
of this unconventional superconducting state is
finely related to the interplay between antiferromagnetism and superconductivity,
having a spatially inhomogeneous superconducting order
maximal when the two parent orders coexist.
We finally showed that this phenomenology happens for generic quasiperiodic
modulation frequencies and survives the presence of perturbations to the
Hamiltonian.
Ultimately, the phenomenology presented can be realized in
twisted graphene superlattices with atomically engineered
impurities, and generically on moire patterns
between two-dimensional
antiferromagnets and superconductors.
Our results put forward antiferromagnetic-superconducting
quasiperiodicity as a powerful knob to engineer
robust superconducting states, 
providing a new route towards the design of
artificial unconventional superconductors.

\section*{Acknowledgements}
We acknowledge the computational resources provided by
the Aalto Science-IT project. We thank P. Liljeroth,
O. Zilberberg, A. Strkalj, M. Sigrist and A. Ramires
for useful discussions. 
J.L.L. is grateful for financial support from the
Academy of Finland Projects No. 331342 and No. 336243.

\bibliography{myref}
\end{document}